\documentclass[aps,graphicx]{revtex4}
\usepackage{amsmath}
\usepackage{graphicx}

\begin{document}

\title{Universal quantum gates for photon-atom hybrid systems assisted by
bad cavities\footnote{Published in Sci. Rep. \textbf{6}, 24183
(2016)}}

\author{Guan-Yu Wang$^{1}$, Qian Liu$^{1}$, Hai-Rui Wei$^{2}$, Tao Li$^{3}$,
Qing Ai$^{1}$, and Fu-Guo Deng$^{1}$\footnote{Corresponding author: fgdeng@bnu.edu.cn}}

\affiliation{$^{1}$Department of Physics, Applied Optics Beijing
Area Major Laboratory, Beijing normal University, Beijing 100875, China\\
$^{2}$School of Mathematics and Physics, University of Science and
Technology Beijing, Beijing 100083, China\\
$^{3}$State key Laboratory of Low-Dimensional Quantum Physics and
Department of Physics,
 Tsinghua University, Beijing 100084, China}

\date{\today }

\begin{abstract}
We present two deterministic schemes for constructing a  CNOT gate
and a Toffoli gate on photon-atom and photon-atom-atom hybrid
quantum systems assisted by bad cavities, respectively. They are
achieved by cavity-assisted photon scattering and work in the
intermediate coupling region with  bad cavities, which relaxes the
difficulty of their implementation in experiment. Also, bad cavities
are feasible for fast quantum operations and reading out
information. Compared with previous works, our schemes do not need
any auxiliary qubits and measurements. Moreover, the schematic
setups for these gates are simple, especially that for our Toffoli
gate as  only a quarter wave packet is used to interact the photon
with each of the atoms every time. These atom-cavity systems can be
used as the quantum nodes in long-distance quantum communication as
their relatively long coherence time is suitable for multi-time
operations between the photon and the system. Our calculations show
that the average fidelities and efficiencies of our two universal
hybrid quantum gates are high with current experimental technology.
\end{abstract}\maketitle

A quantum computer \cite{quantum1} can run the famous Shor's
algorithm \cite{c6} for integer factorization and implement
Grover-Long algorithm \cite{Grover,GroverLong} for unsorted database
search. In past decades, it has attracted much attention. Quantum
logic gates are the key elements in  quantum computers and  play a
critical role in quantum information processing (QIP). Two-qubit
controlled-not (CNOT) gates together with single-qubit gates are
sufficient for universal quantum computing \cite{quantum1,quantum2}.
In 2004, Shenge proposed a ``small-circuit'' structure which is used
to construct CNOT gates \cite{c3}. In the domain of three-qubit
gates, Toffoli gate has attracted much attention and it is
universal. Together with Hadamard gates, it can realize unitary
manipulation for a multi-qubit system \cite{c4,c5}. Moreover, it
plays an important role in phase estimation \cite{quantum1}, complex
quantum algorithms \cite{c6,Grover,GroverLong}, error correction
\cite{c7}, and fault tolerant quantum circuits \cite{c8}. In 2009,
the optimal synthesis for a Toffoli gate with six CNOT gates was
proposed \cite{c9}. Up to now,  for a general three-qubit logic
gate, the optimal synthesis requires twenty CNOT gates \cite{c10},
which means that this method increases the difficulty and complexity
of experiments and the possibility of errors largely. It is
significant to seek a simpler scheme to directly implement
multi-qubit gates.

By far, many physical systems have been used to implement quantum
logic gates, such as photons in the polarization degree of freedom
(DOF) \cite{photon1,photon2,photon3} and those in both the
polarization and the spatial-mode DOFs (the hyper-parallel photonic
quantum computing) \cite{HyperCNOT1,HyperCNOT2,Ren}, nuclear
magnetic resonance \cite{NMR1,NMR2,NMR3,NMR4}, quantum dots
\cite{spin1,spin2,spin3,spin4,spin5}, diamond nitrogen-vacancy
center \cite{NV1,NV2,NV3}, superconduting qubits
\cite{superconducting1,superconducting2}, superconducting resonators
(microwave photons) \cite{microwave1,microwave2}, and hybrid quantum
systems \cite{hybrid,Wei}. Cavity quantum electrodynamics (QED) is a
promising physical platform  for constructing universal quantum
logic gates as it can enhance the interaction between a photon and
an atom (or an artificial atom). Because of the robustness against
decoherence,  photons are the perfect candidates for fast and
reliable long-distance communication. Meanwhile, the stationary
qubits are suitable for processor and local storage. Quantum logic
gates between flying photon qubits and stationary qubits hold a
great promise for quantum communication and computing, especially
for quantum repeaters, distributed quantum computing, and blind
quantum computing. In 2013, Wei and Deng \cite{Wei} proposed some
interesting schemes for universal hybrid quantum gates which use
quantum dots inside double-sided optical microcavities as stationary
qubits and the flying photon as the control qubit.  An atom trapped
in an optical microcavity is an attractive candidate for a
stationary qubit. The interaction time  between a single atom and
the cavity in which the atom is trapped can be maintained for 10 s
\cite{c20}. By using the atoms interacting with local cavities as
the quantum nodes and the photon transmitting between remote nodes
as the quantum bus, one can set up a quantum network to realize a
large-scale QIP.

Many schemes \cite{c11,c12,c13,c14,c15,c16,c21,c26,c28} for QIP
tasks, assisted by the input-output process in atom-cavity system,
have been proposed. In 2004, Duan and Kimble \cite{c14} proposed a
scheme for the construction of a controlled phase-flip (CPF) gate
between an atom trapped in a cavity and a single photon. The strong
coupling between the atom and the cavity can provide a large Kerr
nonlinearity. Combined with the input-output process of the flying
single photon, a universal quantum gate can be achieved \cite{c14}.
Interestingly, the atom-photon coupling in a optical cavity have
been implemented in experiments. For example, Reiserer \emph{et al.}
\cite{Reisterer1} demonstrated an optical nondestructive detection
based on reflecting a photon from an optical cavity \cite{c14}
containing a single atom. In 2014, Tiecke \emph{et al.}
\cite{Tiecke} realized a system in which a single atom, trapped in a
photonic crystal cavity, switches the phase of a photon and a single
photon modifies the phase of an atom. In the same year, Reiserer
\emph{et al.} \cite{Reisterer2} implemented a CPF gate between the
spin state of a single trapped atom and the polarization state of a
photon. In 2015, Kalb \emph{et al.} \cite{Kalb} realized a heralded
transfer of a polarization qubit from a photon onto a single atom.
It is significant for seeking a realization of QIP task in the weak
coupling region with a bad cavity. In 1995, Turchette \emph{et al.}
\cite{Turchette} completed a measurement of conditional phase shifts
for quantum logic in an intermediate atom-cavity coupling regime
with a bad cavity. In 2008, Dayan \emph{et al.} \cite{c30} achieved
an experiment in which the transport of photons is regulated by one
atom trapped in a cavity in an intermediate atom-cavity coupling
regime with a bad cavity. Without the requirements of good cavities
or strict strong coupling strength, many theoretical QIP tasks have
been proposed, such as quantum gates
\cite{c25,An,An6,An5,An1,An4,An3}, generation of entangled states
\cite{An7}, and quantum controlled teleportation \cite{An2}. In
2006, Xiao \emph{et al.} \cite{c25} proposed a scheme of CPF gate
without strict strong coupling on a silicon chip. In 2009, An
\emph{et al.} \cite{An} presented a scheme for QIP with a single
photon by an input-output process with respect to bad cavities.
2009, Chen \emph{et al.} \cite{An6} achieved CPF gates by modifying
the original idea proposed by An \emph{et al.} \cite{An}.

In this paper, we present a deterministic scheme for constructing a
CNOT gate on a hybrid photon-atom system through the atom-cavity
photon scattering. In our scheme, the control qubit is encoded on a
flying photon  (i.e., the two polarization states of a single
photon, the right circular polarization and the left circular
polarization), while the target qubit is encoded on the ground
states of an atom trapped in a bad optical microcavity. We also
present a deterministic scheme for constructing a Toffoli gate on a
photon-atom-atom hybrid system. We use the atom-cavity systems as
our quantum nodes to realize our two quantum gates. The long
coherence time of the system is feasible for multi-time operations
between the photon and the system and it is suitable for perfect
quantum memory. These two gates work in the intermediate coupling
region with  bad cavities, not require strong coupling strength with
good cavities, which relaxes the difficulty of their implementation
in experiment. In the bad cavity limit, $\kappa \gg g^{2}/\kappa \gg
\gamma$, it is feasible for fast reading out information, and it is
effective for reducing the interaction time between the photon and
the atom-cavity system. Our two gates do not require any auxiliary
qubits and measurements. Moreover, our schematic setup of the
Toffoli gate is very simple, as only a quarter wave packet is used
to interact the photon with the atom-cavity system every time, which
can reduce the imperfection of the nonlinear interaction. It will be
shown that high average fidelities and efficiencies can be achieved
for these gates with the intermediate coupling between the atom and
the cavity region.

\begin{figure}[htbp]             
\centering
\includegraphics[width=7 cm]{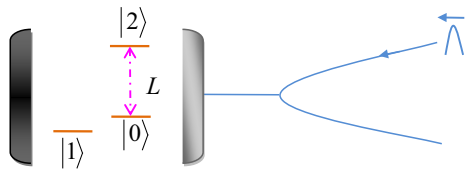}
\caption{ The optical transitions of an atom trapped in a
single-sided optical cavity with circularly polarized lights. The
\emph{left} wall of the cavity is perfectly reflective and the
\emph{right} one is partially reflective. $|0\rangle$, $|1\rangle$,
and $|2\rangle$ represent the two ground states and the one excited
state of the atom, respectively. $R$ ($L$) represents the right
(left) circularly polarized photon.}\label{fig1}
\end{figure}

\section*{Results}

\textbf{The single-photon input-output process.}   Let us consider
an atom which has two ground states $|0\rangle$ and $|1\rangle$ and
an excited state $|2\rangle$ trapped in a single-sided optical
cavity, shown in Fig. 1. The cavity considered here is one side wall
perfectly reflective and the other side wall partially reflective
\cite{c14}. The left-circularly $L$ polarized cavity mode couples
the transition $|0\rangle\leftrightarrow|2\rangle$ (for example, the
D2 transition $(6S_{1/2},F=4,m=4)\rightarrow (6P_{3/2},F'=5,m'=5)$
of cesium), while it decouples the transition
$|1\rangle\leftrightarrow|2\rangle$ because of large detuning. Under
the Jaynes-Commings model, the Hamiltonian of the whole system
composed of a single  cavity mode ($L$ polarized) and an atom
trapped in a single-sided cavity can be expressed as:
\begin{eqnarray}    
H\;=\;\frac{\hbar\omega_{0}}{2}\sigma_{z}\,+\,\hbar\omega_{c}a^{\dag}a\,+\,
ig\hbar\left(a\sigma_{+}-a^{\dag}\sigma_{-}\right).
\end{eqnarray}
Here $a$ and $a^{\dagger}$ are the annihilation and creation
operators of the $L$ polarized cavity mode with the frequency
$\omega_{c}$, respectively. $\sigma_{z}$, $\sigma_{+}$, and
$\sigma_{-}$ are the inversion, raising, and lowering operators of
the atom, respectively. $\omega_{0}$ is the frequency difference
between the ground level $|0\rangle$ and the excited level
$|2\rangle$ of the atom. $g$ is the atom-cavity coupling strength,
which is affected by the trapping position of the atom. The
reflection coefficient of a single-photon pulse with the frequency
$\omega_{p}$ injected into the optical cavity can be obtained by
solving the Heisenberg- Langevin equations of  motion for the
internal cavity field and the atomic operators in the interaction
picture \cite{c32}:
\begin{eqnarray}    \label{eq2}
\begin{split}
\dot{a}(t)&\;=\;-\left[i(\omega_{c}-\omega_{p})+\frac{\kappa}{2}\right]a(t)-g\sigma_{-}(t)-\sqrt{\kappa}\,a_{in}(t),    \\
\dot{\sigma_{-}}(t)&\;=\;-\left[i(\omega_{0}-\omega_{p})+\frac{\gamma}{2}\right]\sigma_{-}(t)-g\,\sigma_{z}(t)\,a(t) +\sqrt{\gamma}\,\sigma_{z}(t)\,b_{in}(t), \\
 a_{out}(t)&\;=\;a_{in}(t)+\sqrt{\kappa}\,a(t).
\end{split}
\end{eqnarray}
Here the one-dimensional field operator $a_{in}(t)$ is the cavity
input operator which satisfies the commutation relation
$[a_{in}(t),a^{\dagger}_{in}(t')]=\delta(t-t')$. $b_{in}(t)$, with
the commutation relation
$[b_{in}(t),b^{\dagger}_{in}(t')]=\delta(t-t')$, is the vacuum input
field felt by the three-level atom. $a_{out}$ is the output
operator. $\kappa$ and $\gamma$ are the cavity damping rate and the
atomic decay rate, respectively.

The atom is prepared in the ground states initially. By making
$\kappa$ sufficiently large, one can ensure that the excitation by a
single-photon pulse is a weak one, and  obtain the input-output
relation of the cavity field \cite{An}
\begin{eqnarray}  \label{eq3}  
r(\omega_{p})\;=\;\frac{\left[i\left(\omega_{c}-\omega_{p}\right)-\frac{\kappa}{2}\right]\left[i\left(\omega_{0}-\omega_{p}\right)+\frac{\gamma}{2}\right]
+g^{2}}{\left[i\left(\omega_{c}-\omega_{p}\right)+\frac{\kappa}{2}\right]\left[i\left(\omega_{0}-\omega_{p}\right)+\frac{\gamma}{2}\right]+g^{2}}.
\end{eqnarray}
Here $r(\omega_{p})\equiv\frac{a_{out}(t)}{a_{in}(t)}$ is the
reflection coefficient for the atom-cavity system. When the atom is
uncoupled to the cavity or an empty cavity, that is, $g=0$, one can
obtain \cite{c32}
\begin{eqnarray}   \label{eq4}  
r_{_0}(\omega_{p})\;=\;\frac{i(\omega_{c}-\omega_{p})-\frac{\kappa}{2}}{i(\omega_{c}-\omega_{p})+\frac{\kappa}{2}}.
\end{eqnarray}

If the atom is initially prepared in the ground state $|0\rangle$,
the left-circularly polarized single-photon pulse $|L\rangle$ will
drive the transition $|0\rangle\leftrightarrow|2\rangle$. The output
pulse related to the input one can be expressed as
$|\Phi_{out}\rangle_{L}=r(\omega_{p})|L\rangle\approx
e^{i\phi}|L\rangle$. The phase shift $\phi$ is determined by the
parameter values in Eq.(\ref{eq3}). However, if the atom is
initially prepared in the ground state $|1\rangle$, the
left-circularly polarized single-photon $|L\rangle$ will only sense
a bare cavity. As a result, the corresponding output governed by
Eq.(\ref{eq4}) is
$|\Phi_{out}\rangle_{L}=r_{_0}(\omega_{p})|L\rangle\approx
e^{i\phi_{_0}}|L\rangle$, with a phase shift $\phi_{_0}$ different
from $\phi$. Considering the parameters of the atom-cavity system
satisfy the relationship $\omega_{0}=\omega_{c}=\omega_{p}$, the
reflection coefficient can be expressed as
\begin{equation}\label{5}
\begin{split}
r(\omega_{p})
\;=\;\frac{-1+(2g/\sqrt{\kappa\gamma})^{2}}{1+(2g/\sqrt{\kappa\gamma})^{2}},\;\;\;\;\;\;\;\;\;\;\;\;\;\;\;\;\;\;\;
r_{_0}(\omega_{p}) \;=\;-1.
\end{split}
\end{equation}
Considering a bad cavity $\kappa \gg g^{2}/\kappa\gg \gamma$ in the
atom-cavity intermediate coupling region, phase shifts $\phi=0$ and
$\phi_{0}=\pi$ from Eq. (\ref{5}) can be produced.

\begin{figure}[!h]
\begin{center}
\includegraphics[width=7 cm,angle=0]{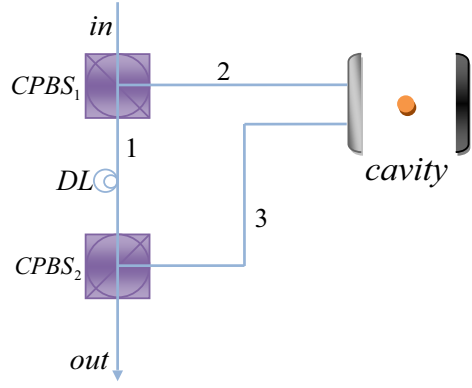}
\caption{ Schematic setup for a deterministic CNOT gate with  a
flying polarized photon as the control qubit and an atom trapped in
a single-sided optical cavity as the target qubit. $CPBS_{i}$
($i=1,2$) is a circularly polarizing beam splitter which transmits
the photon in the right-circular polarization $|R\rangle$ and
reflects the photon in the left-circular polarization $|L\rangle$,
respectively. $M$ is a mirror. $DL$ is a time-delay device which
makes the two wavepackets coming from the paths 2 and 3 interfere
with each other.}\label{fig2}
\end{center}
\end{figure}

{\bf CNOT gate on a two-qubit hybrid system.} Our  CNOT  gate  on  a
two-qubit hybrid system is used to complete a bit-flip on the atom
trapped in the cavity when the flying photon is in the left-circular
polarization $|L\rangle$; otherwise, it does nothing.  The schematic
setup for our CNOT gate is shown in Fig. 2. We will describe its
principle in detail as follows.

Suppose that the initial states of the flying photon $p$ and the
atom $a$ trapped in the single-sided cavity are
\begin{eqnarray}\label{8}
\begin{split}
|\psi_{p}\rangle&\;=\;\alpha_{p}|R\rangle+\beta_{p}|L\rangle, \\
|\psi_{a}\rangle&\;=\;\alpha|0\rangle+\beta|1\rangle.
\end{split}
\end{eqnarray}
First, the flying photon is led to the device shown in Fig. 2. The
circularly polarizing beam splitter $CPBS_{1}$ transmits the  photon
in the right-circular polarization $|R\rangle$ to path $1$ and
reflects the photon in the left-circular polarization $|L\rangle$ to
path $2$. The state of the hybrid system composed of the flying
photon $p$ and the atom $a$ is changed from $|\Psi\rangle_{0} \equiv
|\psi_{p}\rangle \otimes |\psi_{a}\rangle$ to $|\Psi\rangle_{1}$.
Here
\begin{eqnarray}\label{9}
|\Psi\rangle_{1}\;=\;\alpha_{p}|R\rangle_1(\alpha|0\rangle+\beta|1\rangle)_a
+\beta_{P}|L\rangle_2(\alpha|0\rangle+\beta|1\rangle)_a,
\end{eqnarray}
where the subscripts 1 and 2 represent the paths that the flying
photon passes through. The subscript $a$  represents the atom
trapped in the cavity.

Second, a Hadamard operation is performed on the atom trapped in the
cavity before the photon interacts with the atom-cavity system. The
Hadamard operation on the atom is used to complete the
transformations $|0\rangle \rightarrow
\frac{1}{\sqrt{2}}(|0\rangle+|1\rangle)$ and $|1\rangle \rightarrow
\frac{1}{\sqrt{2}}(|0\rangle-|1\rangle)$. Thus, the state of the
hybrid system is changed to be
\begin{eqnarray}\label{10}
\begin{split}
|\Psi\rangle_{2}\;=\;\frac{1}{\sqrt{2}}\Big\{\alpha_{p}|R\rangle_1\big[\alpha(|0\rangle+|1\rangle)+\beta(|0\rangle-|1\rangle)\big]_a
+
\beta_{p}|L\rangle_2\big[\alpha(|0\rangle+|1\rangle)+\beta(|0\rangle-|1\rangle)\big]_a\Big\}.
\end{split}
\end{eqnarray}

Third, the photon interacts with the atom trapped in the
single-sided cavity and the state of the system becomes
\begin{eqnarray}\label{11}
\begin{split}
|\Psi\rangle_{3}\;=\;
\frac{1}{\sqrt{2}}\Big\{\alpha_{p}|R\rangle_1\big[\alpha(|0\rangle+|1\rangle)+\beta(|0\rangle-|1\rangle)\big]_a
+
\beta_{p}|L\rangle_3\big[\alpha(|0\rangle-|1\rangle)+\beta(|0\rangle+|1\rangle)\big]_a\Big\}.
\end{split}
\end{eqnarray}

After the interaction between the flying photon and the atom trapped
in the cavity, a Hadamard operation is performed on the atom again.
At last, the two wavepacks split by $CPBS_{1}$ reunion at $CPBS_{2}$
from path $1$ and path $3$. The state of the system is transformed
into
\begin{equation}\label{12}
|\Psi\rangle_{4}\;=\;\alpha_{p}|R\rangle(\alpha|0\rangle+\beta|1\rangle)_{a}+\beta_{p}|L\rangle(\alpha|1\rangle+\beta|0\rangle)_{a}.
\end{equation}
Here $|\Psi\rangle_{4}$ is the objective state. One can see that the
state of the atom (the target qubit) is flipped when  the photon
(the control qubit) is in the left-circular polarization
$|L\rangle$; otherwise, nothing is done on the atom. That is, the
schematic setup shown in Fig. 2 can be used to deterministically
achieve a quantum CNOT gate on the photon-atom hybrid system by
using the flying photon as the control qubit and the atom as the
target qubit in principle.

\begin{figure}[!h]
\begin{center}
\includegraphics[width=10 cm,angle=0]{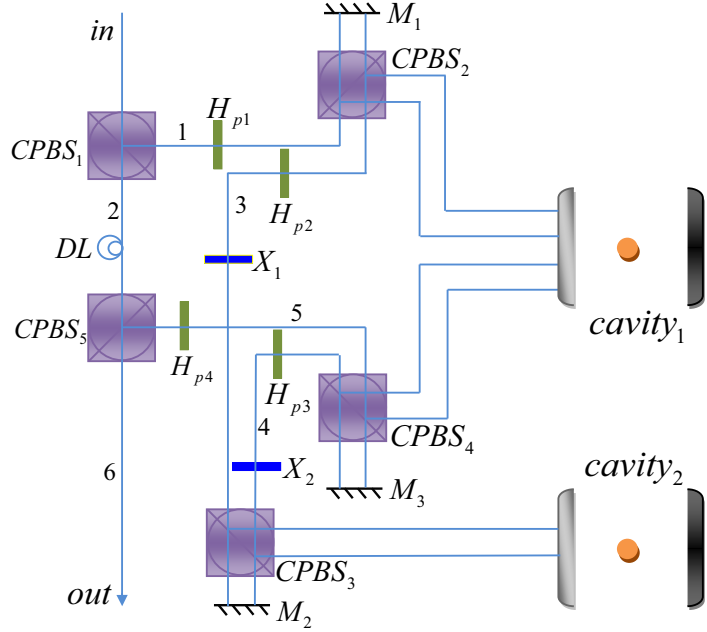}
\caption{Schematic setup for constructing a deterministic Toffoli
gate with the polarization of a flying photon and an atom trapped in
a single-sided cavity ($cavity_1$) as the two control qubits and
anther atom trapped in another single-sided cavity ($cavity_2$) as
the target qubit. $H_{pi}(i=1,2,3,4)$ is a half-wave plate with the
axis at $22.5^{\circ}$ and it performs a Hadamard operation on the
photon. $M_{i}$ ($i=1,2,3,4,5$) is a mirror. $cavity_{i}$ ($i=1,2$)
represents the atom-cavity system. }\label{fig3}
\end{center}
\end{figure}

\textbf{Toffoli gate on a three-qubit hybrid system.}  Our Toffoli
gate on a three-qubit hybrid system is used to complete a bit-flip
operation on the atom trapped in $cavity2$ (the target qubit) when
the polarization of the flying photon (the first control qubit) is
in the left-circular polarization $|L\rangle$ and the atom trapped
in $cavity1$ (the second control qubit) is in the state $|1\rangle$
at the same time; otherwise, it does nothing on the atom trapped in
$cavity2$.  The schematic setup of our Toffoli gate is shown in Fig.
3. Assume that the initial states of the flying photon qubit and the
two atoms trapped in $cavity$ 1 and $cavity$ 2 are prepared in
$|\phi_{p}\rangle$, $|\phi_{a1\rangle}$, and $|\phi_{a2}\rangle$,
respectively. Here,
\begin{eqnarray}\label{13}
\begin{split}
|\phi_{p}\rangle&\;=\;\alpha_{p}|R\rangle+\beta_{p}|L\rangle,   \\
|\phi_{a1}\rangle&\;=\;\alpha_{1}|0\rangle+\beta_{1}|1\rangle,   \\
|\phi_{a2}\rangle&\;=\;\alpha_{2}|0\rangle+\beta_{2}|1\rangle.
\end{split}
\end{eqnarray}
The principle of our Toffoli gate can be described in detail as
follows.

First, the photon is led into our device from the port $in$.
$CPBS_{1}$ reflects the photon in the left-circular polarization
$|L\rangle$ to  path $1$ and transmits the photon in the
right-circular polarization $|R\rangle$ to  path $2$. The photon
passing through path $2$ will not interact with the atoms trapped in
cavities. After the photon passes through $CPBS_{1}$, the state of
the system is changed from $|\Phi\rangle_{0} \equiv |\phi_{p}\rangle
\otimes |\phi_{a1}\rangle \otimes |\phi_{a2}\rangle$ to
$|\Phi\rangle_{1}$. Here,
\begin{eqnarray}\label{14}
\begin{split}
|\Phi\rangle_{1}\;=\;&\alpha_{p}|R\rangle_{2} \otimes
(\alpha_{1}|0\rangle+\beta_{1}|1\rangle)_{a_1}
\otimes (\alpha_{2}|0\rangle+\beta_{2}|1\rangle)_{a_2}  \\
& +\;\beta_{p}|L\rangle_{1} \otimes
(\alpha_{1}|0\rangle+\beta_{1}|1\rangle)_{a_1} \otimes
(\alpha_{2}|0\rangle+\beta_{2}|1\rangle)_{a_2}.
\end{split}
\end{eqnarray}

Second, a Hadamard operation is performed on the photon in path $1$,
and $CPBS_{2}$ transmits the photon in $|R\rangle$ to $M_{1}$ and
reflects the photon in $|L\rangle$ to $cavity$ 1. Here the Hadamard
operation on the photon completes the transformations $|R\rangle
\rightarrow \frac{1}{\sqrt{2}} (|R\rangle +|L\rangle)$ and
$|L\rangle \rightarrow  \frac{1}{\sqrt{2}} ( |R\rangle -|L\rangle)$.
Subsequently,  the flying photon interacts with the atom trapped in
$cavity$ 1. After the interaction, the two components of the photon
reunion at $CPBS2$. Also, a Hadamard operation and a bit-flip
operation $\sigma_{x}=|L\rangle\langle R|+|R\rangle \langle L|$ are
performed on the photon in path $3$. The state of the whole system
becomes
\begin{eqnarray}\label{15}
\begin{split}
|\Phi\rangle_{2}\;=\;&\alpha_{p}|R\rangle_{2}(\alpha_{1}|0\rangle+\beta_{1}|1\rangle)_{a_1}\otimes(\alpha_{2}|0\rangle+\beta_{2}|1\rangle)_{a_2}\\
&
+\;\beta_{p}|R\rangle_{3}\alpha_{1}|0\rangle_{a1}\otimes(\alpha_{2}|0\rangle+\beta_{2}|1\rangle)_{a_2}+\beta_{p}|L\rangle_{3}\beta_{1}|1\rangle_{a1}\otimes(\alpha_{2}|0\rangle+\beta_{2}|1\rangle)_{a_2}.
\end{split}
\end{eqnarray}

Third, one can perform a Hadamard operation on the atom trapped in
$cavity$ 2 and lead the photon in $|L\rangle$ reflected by
$CPBS_{3}$ to $cavity$ 2 and the photon in $|R\rangle$ transmitted
by $CPBS_{3}$ to $M_{2}$. The photon in $|L\rangle$ and the atom
trapped in $cavity$ 2 interact with each other. After the
interaction, a Hadamard operation is performed on the atom trapped
in $cavity$ 2 again. The two components of the photon reunion at
$CPBS_{3}$. The state of the whole system is changed into
\begin{eqnarray}\label{16}
\begin{split}
|\Phi\rangle_{3}\;=\;&\alpha_{p}|R\rangle_{2}(\alpha_{1}|0\rangle
+\beta_{1}|1\rangle)_{a_1}\otimes(\alpha_{2}|0\rangle+\beta_{2}|1\rangle)_{a_2}\\
&
+\;\beta_{p}|R\rangle_{4}\alpha_{1}|0\rangle_{a1}(\alpha_{2}|0\rangle+\beta_{2}|1\rangle)_{a_2}+\beta_{p}|L\rangle_{4}\beta_{1}|1\rangle_{a1}(\alpha_{2}|1\rangle+\beta_{2}|0\rangle)_{a_2}.
\end{split}
\end{eqnarray}

Finally, a bit-flip operation and a Hadamard operation are performed
on the photon which emerges in  path $4$. $CPBS_{4}$ transmits the
photon in $|R\rangle$ to $M_{3}$ and reflects the photon in
$|L\rangle$ to $cavity$ 1. The photon in $|L\rangle$ interacts with
the atom trapped in $cavity$ 1 again. After the interaction between
the atom-cavity system and the photon, $CPBS_{4}$ reflects the
photon in $\vert L\rangle$ and transmits the photon in $\vert
R\rangle$ to path $5$.  The former is reflected by $cavity$ 1 and
the latter is reflected by $M_{3}$.  A Hadamard operation is
performed on the photon in path $5$. At this time, the two
components of the photon from paths $2$ and  $5$ pass through
$CPBS_{5}$ simultaneously, and then the photon is led out of our
device. The final state of the whole system composed of the flying
photon and the two atoms trapped in two cavities separately can be
expressed as
\begin{eqnarray}\label{eq17}
\begin{split}
|\Phi\rangle_{f}\;=\;&\alpha_{p}|R\rangle
\,\alpha_{1}|0\rangle_{a_1}
(\alpha_{2}|0\rangle+\beta_{2}|1\rangle)_{a_2}
 +\beta_{p}|L\rangle \,\alpha_{1}|0\rangle_{a_1} (\alpha_{2}|0\rangle+\beta_{2}|1\rangle)_{a_2}     \\
& +\; \alpha_{p}|R\rangle \,\beta_{1}|1\rangle_{a_1}
(\alpha_{2}|0\rangle+\beta_{2}|1\rangle)_{a_2}     +
\beta_{p}|L\rangle \,\beta_{1}|1\rangle_{a_1}
(\alpha_{2}|1\rangle+\beta_{2}|0\rangle)_{a_2}.
\end{split}
\end{eqnarray}
From Eq. (\ref{eq17}), one can see that the state of the atom
trapped in $cavity$ 2 (the target qubit) is flipped only  when  the
photon (the first control qubit) is in the left-circular
polarization $|L\rangle$ and the atom trapped in $cavity$ 1 (the
second control qubit) is in $|1\rangle$ at the same time. That is,
the schematic setup shown in Fig. 3 can achieve a quantum Toffoli
gate on a photon-atom-atom hybrid system by using the flying photon
and the atom in
 $cavity$ 1 as the two control qubits and the atom in $cavity$ 2
as the target qubit in  a deterministic way.

\section*{Discussion}

In 2013, Reiserer \emph{et al.} \cite{Reisterer1} exploited the
atom-cavity system to complete a robust photon detection scheme
experimentally. In their experiment,  a single $^{87}$Rb atom is
trapped at the center of a Fabry-Perot resonator \cite{Reisterer3}.
Their experiment was completed in the experimental parameters
$[g,\kappa,\gamma]/2\pi=[6.7,2.5,3.0]MHz$. In the same experimental
parameters, they \cite{Reisterer2} implemented a quantum CNOT gate
that a flip of the photon is controlled by an atom trapped in a
Fabry-Perot cavity. In 2014, Tiecke\emph{et al.} \cite{Tiecke}
realized a scheme in which a single atom switches the phase of a
photon and a single photon modifies the atom's phase. Their
experiment was implemented in the parameters
$[2g,\kappa,\gamma]/2\pi=[(1.09\pm0.03)GHz,25GHz,6MHz]$. Compared
with intermediate coupling strength of the atom-cavity system, it is
still  challenging to realize the strong coupling strength in
experiment. For obtaining shorter operation time, it is significant
to realize the atom-cavity photon scattering with a bad cavity in
experiment. In 1995, Turchette \emph{et al.} \cite{Turchette} made a
measurement on the conditional phase shifts for quantum logic in the
experimental parameters $[g,\kappa,\gamma]/2\pi=[20,75,2.5]MHz$.
These parameters satisfy the limit of a bad cavity $\kappa\gg
g^{2}/\kappa \gg \gamma$ and an intermediate coupling region
($g=0.27\kappa$ ). Based on these experimental parameters, the
average fidelities of our CNOT gate and Toffoli gate are
$\overline{F}_{C}=0.9943$ and $\overline{F}_{T}=0.9885$,
respectively. The average efficiencies of our CNOT gate and Toffoli
gate are $\overline{P}_{C}=0.9061$ and $\overline{P}_{T}=0.8631$,
respectively. In 2008, Dayan \emph{et al.} \cite{c30} demonstrated
an intermediate atom-cavity coupling in experiment. In their
experiment, a Cs atom is trapped in a microtoroidal resonator. They
gave a set of parameters
$[g,\kappa,\gamma]/2\pi=[70,(165\pm15),2.6]MHz$. The probe laser can
be swept continuously over a range
$\Delta=\omega_{p}-\omega_{c}=\pm400$ MHz and the atom-cavity
detuning $\omega_{0}-\omega_{c}=0$ can be obtained. The parameters
in their experiment satisfy the requirements of a bad cavity and an
intermediate coupling regime ($g=0.38\kappa$ ). Based on these
experimental parameters, the average fidelities of our CNOT gate and
Toffoli gate are $\overline{F}_{C}=0.9998$ and
$\overline{F}_{T}=0.9994$, respectively. The average efficiencies of
our CNOT gate and Toffoli gate are $\overline{P}_{C}=0.9772$ and
$\overline{P}_{T}=0.9661$, respectively. The analyses above show
that the average fidelities and the averages efficiencies of our two
gates can remain high values in the intermediate coupling region
with a bad cavity.

In contrary to the CNOT scheme presented by Bonant \emph{et al}.
\cite{hybrid}, in which a confined electron spin in a QD trapped in
a cavity acts as a control qubit and the spin of the photon acts as
a target qubit, we use a flying photon as a control qubit and use an
atom trapped in an cavity as a target qubit. Our scheme is different
from the CNOT scheme proposed by Reiserer \emph{et al}.
\cite{Reisterer2}, in which an atom trapped in a cavity acts as a
control qubit and the polarization state of the photon acts as a
target qubit. In our scheme, the two different polarizations of the
photon are split by the CPBS before the photon interacts with the
atom-cavity system, which will reduce the difficulty of the
experiment. Our scheme is also different from the work by Su
\emph{et al.} \cite{An3} in which an atom trapped in an cavity acts
as the control qubit and an atom trapped in another cavity acts as
the target qubit with an auxiliary atom qubit and measurements on
the auxiliary qubit and the photon.

In summary,  we have proposed two schemes for constructing a
deterministic CNOT gate and a deterministic Toffoli gate on
photon-atom hybrid systems, respectively, by utilizing the nonlinear
interaction between the flying photon and the atom-cavity system and
some linear optical elements.  For our CNOT gate, the control qubit
is encoded on the flying  photon and the target qubit is encoded on
the atom trapped in the cavity. For our Toffoli gate, the control
qubits are encoded on the flying photon and an atom trapped in one
cavity and the target qubit is encoded on an atom trapped in another
cavity. The quantum circuits of our two gates are  very simple. They
do not need any auxiliary qubit and measurements to complete the
CNOT  and Toffoli gates on photon-atom hybrid systems. Our two
schemes can work in the atom-cavity intermediate coupling region
with bad cavities. The atom-cavity system working in the
intermediate coupling region is achieved in experiment
\cite{Turchette,c30}. The ratio of coupling strength to dissipation
factors $g/\sqrt{\kappa\gamma}$ affects the fidelities and
efficiencies of our gates a little. Our calculations show that even
in a worst condition or a reasonable experimental condition, the
average of fidelities and the average efficiencies of our two gates
can remain high values. What's more, there exist experimental
parameters that satisfy the requirements in work.

\begin{figure}[!h]
\centering
\includegraphics[width=16cm,angle=0]{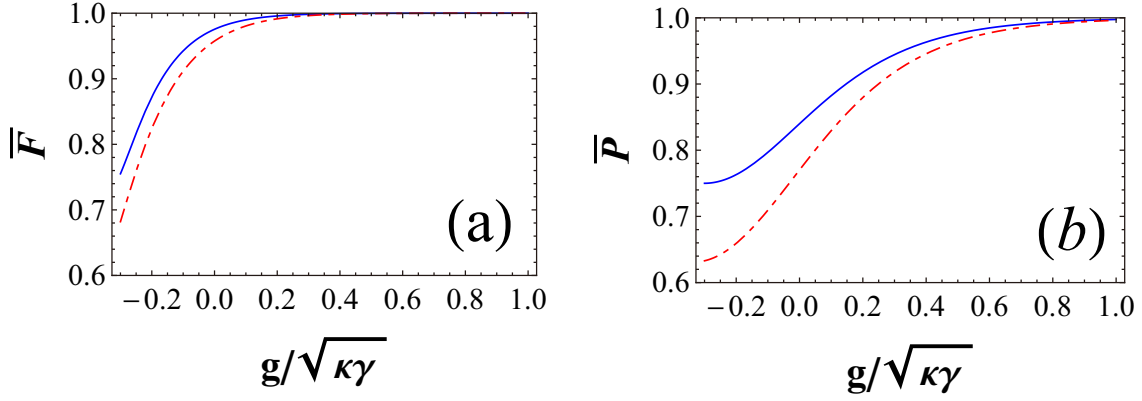}
\caption{ (a) Average fidelity $\overline{F}$ of our CNOT gate on a
two-qubit hybrid system (solid line) and that of our Toffoli gate on
a three-qubit hybrid system (dashed line) vs $\gamma/\kappa$ on a
logarithmic scale. (b) Average efficiency $\overline{P}$ of our CNOT
gate on a two-qubit hybrid system (solid line) and that of our
Toffoli gate on a three-qubit hybrid system (dashed line) vs
$\gamma/\kappa$ on a logarithmic scale.}\label{fig4}
\end{figure}

\section*{Methods}

\textbf{Fidelities and efficiencies of the gates.} The nonlinear
interaction between the single photon and the atom-cavity system
produces a phase shift between the output photon and the input
photon. Utilizing this shift and some linear optical elements we
construct a CNOT gate and a Toffoli gate on photon-atom and
photon-atom-atom hybrid quantum systems, respectively. In the
process of constructing these two universal quantum hybrid gates, we
set $\phi_{_0}=\pi$ and $\phi=0$. In this ideal case, the hybrid
quantum gates are deterministic, and the fidelity and the efficiency
are $100\%$ for each gate. However, the phase shift $\phi_{_0}=\pi$
is an exact value when $g=0$, while the phase shift $\phi=0$ is an
approximate value when $\kappa \gg g^{2}/\kappa \gg \gamma$. It is a
function of $g/\sqrt{\kappa\gamma}$, which is decided by the
experimental condition. Considering the realistic condition, we will
calculate the fidelities of our quantum gates to show their
performance. The fidelity is defined as $
F=|\langle\Psi_{r}|\Psi_{i}\rangle|^{2}$. Here $|\Psi_{r}\rangle$
and $|\Psi_{i}\rangle$ are final states of the hybrid quantum system
in our schemes for quantum gates in the realistic condition and the
ideal condition, respectively.

The fidelity of our CNOT gate is expressed as
\begin{equation}\label{18}
F_{C}\;=\;\frac{|\alpha_{p}^{2}+\frac{1}{2}\beta_{p}^{2}[2\alpha\beta(r-1)+r+1]|^{2}}{|\alpha_{p}|^{2}+\frac{1}{4}|\alpha\beta_{p}(r-1)+\beta\beta_{p}(r+1)|^{2}+\frac{1}{4}|\alpha\beta_{p}(r+1)+\beta\beta_{p}(r-1)|^{2}}.
\end{equation}
The coefficients of the system can be expressed as
$\alpha_{p}=\cos\varphi$, $\beta_{p}=\sin\varphi$,
$\alpha=\cos\theta$, and $\beta=\sin\theta$. The average fidelity of
the CNOT gate is
\begin{equation}\label{19}
\overline{F}_{C}\;=\;\frac{1}{4\pi^{2}}\int_{0}^{2\pi}\int_{0}^{2\pi}F_{C}\,d\varphi\
d\theta.
\end{equation}
The relationship between the average fidelity of our CNOT gate and
$g/\sqrt{\kappa\gamma}$ on a logarithmic scale is shown in Fig. 4(a)
with the solid line. For our Toffoli gate on a three-qubit hybrid
system, its fidelity is
\begin{equation}\label{20}
F_{T}=\frac{|\alpha_{p}^{2}+\frac{\beta_{p}^{2}\alpha_{1}^{2}}{8}[(r-1)^{3}+2\alpha_{2}\beta_{2}(r-1)^{2}(r+1)+2(r+1)^{2}]+\frac{\beta_{p}^{2}\beta_{1}^{2}}{2}[2\alpha_{2}\beta_{2}(r-1)+r+1]|^{2}}{\xi_{1}+\xi_{2}+\xi_{3}+\xi_{4}+\xi_{5}+\xi_{6}}.
\end{equation}
where $\xi_{1}=|\alpha_{p}|^{2}$,
$\xi_{2}=\frac{1}{64}|\beta_{p}\alpha_{1}[\alpha_{2}(r-1)^{3}+\beta_{2}(r-1)^{2}(r+1)+2\alpha_{2}(r+1)^{2}]|^{2}$,
$\xi_{3}=\frac{1}{64}|\beta_{p}\alpha_{1}[\beta_{2}(r-1)^{3}+\alpha_{2}(r-1)^{2}(r+1)+2\beta_{2}(r+1)^{2}]|^{2}$,
$\xi_{4}=\frac{1}{4}|\beta_{p}\beta_{1}[\alpha_{2}(r-1)+\beta_{2}(r+1)]|^{2}$,
$\xi_{5}=\frac{1}{4}|\beta_{p}\beta_{1}[\alpha_{2}(r+1)+\beta_{2}(r-1)]|^{2}$,
and
$\xi_{6}=\frac{1}{32}|\beta_{p}\alpha_{1}[\alpha_{2}(1+r)^2(1-r)+\beta_{2}(1+r)^2(1-r)]|^{2}$.
The coefficients of the system can be expressed as
$\alpha_{p}=\cos\varphi$, $\beta_{p}=\sin\varphi$,
$\alpha_{1}=\cos\theta$, $\beta_{1}=\sin\theta$,
$\alpha_{2}=\cos\eta$, and $\beta_{2}=\sin\eta$. The average
fidelity of our Toffoli gate is
\begin{equation}\label{21}
\overline{F}_{T}=\frac{1}{8\pi^{3}}\int_{0}^{2\pi}\int_{0}^{2\pi}\int_{0}^{2\pi}F_{T}\,d\varphi
d\theta d\eta.
\end{equation}
The dashed line in Fig. 4(a) shows the relationship between the
average fidelity of  our Toffoli gate and $g/\sqrt{\kappa\gamma}$ on
a logarithmic scale.

The efficiency of a quantum gate is defined as
$P=\frac{n_{out}}{n_{in}}$, where $n_{out}$ is the number of the
photons coming out of the device and $n_{in}$ is the number of the
photons led into the device.  The efficiency of our CNOT gate is
\begin{equation}\label{22}
P_{\;C}=|\alpha_{p}|^{2}+\frac{1}{4}|\alpha\beta_{p}(r-1)+\beta\beta_{p}(r+1)|^{2}+\frac{1}{4}|\alpha\beta_{p}(r+1)+\beta\beta_{p}(r-1)|^{2}.
\end{equation}
The average efficiency of our CNOT gate is
\begin{equation}\label{23}
\overline{P}_{C}=\frac{1}{4\pi^{2}}\int_{0}^{2\pi}\int_{0}^{2\pi}P_{\;C}d\varphi
d\theta.
\end{equation}
The relationship between the average efficiency of our CNOT gate and
$g/\sqrt{\kappa\gamma}$ on a logarithmic scale is shown in Fig. 4(b)
with the  solid line.

The efficiency of our Toffoli gate is
\begin{equation}\label{24}
P_{\;T}=\xi_{1}+\xi_{2}+\xi_{3}+\xi_{4}+\xi_{5}.
\end{equation}
We can also calculate the average efficiency of our Toffoli gate
\begin{equation}\label{25}
\overline{P}_{T}=\frac{1}{8\pi^{3}}\int_{0}^{2\pi}\int_{0}^{2\pi}\int_{0}^{2\pi}P_{\;T}d\varphi
d\theta d\eta.
\end{equation}
The dashed line in Fig. 4(b) shows the relationship between the
average efficiency of our Toffoli gate and $g/\sqrt{\kappa\gamma}$
on logarithmic scale. From Fig. 4(a) and  (b), one can see that the
average fidelities and average efficiencies of these two universal
quantum gates are affected by the cooperativity $C$ ($\propto
g/\sqrt{\kappa\gamma}$) of the atom-cavity system. The average
fidelities are relatively sensitive to the cooperativity when
$g/\sqrt{\kappa\gamma}<1$ ($0$ on a logarithmic scale) and they are
faintly affected by cooperativity when $g/\sqrt{\kappa\gamma}>1.5$
($0.17$ on a logarithmic scale). If $g/\sqrt{\kappa\gamma}\geq1.5$,
which is not a difficult experimental requirement, the average
fidelities of our CNOT and Toffoli gates can be higher than $0.9949$
and $0.9896$, respectively. The average efficiencies are relatively
sensitive to the cooperativity when $g/\sqrt{\kappa\gamma}<2$ ($0.3$
on a logarithmic scale) and they are faintly affected by the
cooperativity when $g/\sqrt{\kappa\gamma}>3$ ($0.48$ on a
logarithmic scale). If $g/\sqrt{\kappa\gamma}\geq 3$, which is not a
difficult experimental requirement, the average efficiencies of our
CNOT and Toffoli gates can be higher than $0.9737$ and $0.9609$,
respectively.

Except for the cooperativity $C$, some other realistic losses and
imperfections, that would affect the fidelities and the efficiencies
of our schemes, should be taken into account. The mismatching of
spatial mode between cavity and the input photon, the quality of
atomic state preparation and rotation will affect both of the
fidelities and the efficiencies of our schemes \cite{Reisterer2}.
The fidelities of our schemes will be also affected by the small
probability of more than one photon in the input laser pulses
\cite{Reisterer2}. The efficiencies can be also affected by the
stability of difference between the cavity resonance and the
frequency of the input photon and the imperfect absorption losses of
the mirror of the cavity \cite{Reisterer2}. In our scheme, as only
one polarization of a photon is injected to the atom-cavity system
and the two polarizations are split by the $CPBS$, the precise
timing of the arrival times from different photon paths is required
in the realistic experiment.


\textbf{Acknowledgement}: This work is supported by the National
Natural Science Foundation of China under Grant Nos. 11474026  and
11505007, the Fundamental Research Funds for the Central
Universities under Grant No. 2015KJJCA01,  the Youth Scholars
Program of Beijing Normal University under Grant No. 2014NT28, and
the China Postdoctoral Science Foundation under Grant No.
2015M571011.\\

\textbf{Author contributions}: G.Y., Q.L., H.R., Q.A., and F.G.
wrote the main manuscript text, and prepared Figures 1-5. G.Y.,
H.R., T.L. and F.G. did the calculations. F.G. supervised the whole
project. All authors reviewed the manuscript.\\

\textbf{Additional Information}: Competing financial interests: The
authors declare no competing financial interests.


\begin{thebibliography}{}


\bibitem{quantum1} Nielsen, M. A. \& Chuang, I. L. Quantum Computation
and Quantum Information (Cambridge University, Cambridge, 2000).








\bibitem{c6} Shor, P. W. Polynomial-time algorithms for prime factorization and discrete logarithms on a quantum computer.
\emph{SIAM J. Comput.} \textbf{26}, 1484--1509 (1997).


\bibitem{Grover} Grover, L. K. Quantum mechanics helps in searching for a needle in a haystack. \emph{Phys. Rev. Lett.} \textbf{79}, 325 (1997).



\bibitem{GroverLong} Long, G. L. Grover algorithm with zero theoretical failure rate. \emph{Phys. Rev. A} \textbf{64}, 022307 (2001).




\bibitem{quantum2} Barenco, A. \emph{et al.}  Elementary gates for quantum computation.
\emph{Phys. Rev. A} \textbf{52}, 3457--3467 (1995).


\bibitem{c3} Shende, V. V., Bullock, S. S. \& Markov,
I. L. Recognizing small-circuit structure in two-qubit operators.
\emph{Phys. Rev. A} \textbf{70}, 012310 (2005).


\bibitem{c4} Shi, Y. Y. Both Toffoli and controlled-NOT need little help to do universal quantum computing.
\emph{Quant. Inf. Comput.} \textbf{3}, 84--92 (2003).


\bibitem{c5} Fredkin, E. \& Toffoli, T. Conservative logic. \emph{Int. J. Theor. Phys.} \textbf{21}, 219--253 (1982).



\bibitem{c7} Cory, D. G. \emph{et al.} Experimental quantum error correction. \emph{Phys. Rev. Lett.}
\textbf{81}, 2152--2155 (1998).


\bibitem{c8} Dennis, E. Toward fault-tolerant quantum computation without concatenation. \emph{Phys. Rev. A} \textbf{63}, 052314 (2001).


\bibitem{c9} Shende, V. V. \&  Markov, I. L. On the CNOT-cost of Toffoli gates. \emph{Quant. Inf. Comput.} \textbf{9}, 461--486 (2009).


\bibitem{c10} Shende, V. V., Bullock, S. S. \& Markov, I. L. Synthesis of quantum-logic circuits.
\emph{IEEE Trans. on Computer-Aided Design} \textbf{25}, 1000--1010
(2006).



\bibitem{photon1}  Knill, E., Laflamme, R. \& Milburn, G. J. A scheme for efficient quantum computation with linear optics.
\emph{Nature} \textbf{409}, 46--52 (2001).


\bibitem{photon2} O'Brien, J. L., Pryde, G.J., White, A. G., Ralph, T. C. \& Branning, D. Demonstration of an all-optical quantum controlled-NOT gate.
\emph{Nature}  \textbf{426}, 264--267 (2003).


\bibitem{photon3} Wei H. R. \& Deng, F. G. Scalable photonic quantum computing assisted by quantum-dot spin in
double-sided optical microcavity.  \emph{Opt. Express} \textbf{21},
17671--17685 (2013).






\bibitem{HyperCNOT1} Ren, B. C., Wei, H. R. \& Deng, F. G. Deterministic photonic spatial-polarization
hyper-controlled-not gate assisted by a quantum dot inside a
one-side optical microcavity.  \emph{Laser Phys. Lett.} \textbf{10},
095202 (2013).



\bibitem{HyperCNOT2} Ren, B. C. \& Deng, F. G. Hyper-parallel photonic quantum
computation with coupled quantum dots. \emph{Sci. Rep.} \textbf{4},
4623 (2014).


\bibitem{Ren} Ren, B. C., Wang, G. Y. \& Deng, F. G. Universal hyperparallel hybrid photonic quantum gates with
dipole-induced transparency in the weak-coupling regime. \emph{Phys.
Rev. A} \textbf{91}, 032328 (2015).




\bibitem{NMR1} Gershenfeld, N. A. \& Chuang, I. L. Bulk spin-resonance quantum
computation. \emph{Science} \textbf{275}, 350--356 (1997).


\bibitem{NMR2} Jones, J. A., Mosca, M. \& Hansen, R. H. Implementation of a
quantum search algorithm on a quantum computer. \emph{Nature}
\textbf{393}, 344--346 (1998).


\bibitem{NMR3} Long, G. L. \& Xiao, L. Experimental realization of a fetching
algorithm in a 7-qubit NMR spin Liouville space computer.  \emph{J.
Chem. Phys.} \textbf{119}, 8473--8481 (2003).


\bibitem{NMR4} Feng, G. R., Xu, G. F. \& Long, G. L. Experimental realization of
nonadiabatic holonomic quantum computation. \emph{Phys. Rev. Lett.}
\textbf{110}, 190501 (2013).





\bibitem{spin1} Li, X. Q. \emph{et al.} An all-optical quantum gate in a semiconductor
quantum dot. \emph{Science} \textbf{301}, 809--811 (2003).



\bibitem{spin2} Hu, C. Y., Young, A., O'Brien, J. L., Munro, W. J. \& Rarity, J. G. Giant optical Faraday rotation induced by a single-electron spin
in a quantum dot: applications to entangling remote spins via a
single photon. \emph{Phys. Rev. B }\textbf{78}, 085307 (2008).



\bibitem{spin3} Hu, C. Y., Munro, W. J. \& Rarity, J. G. Deterministic photon
entangler using a charged quantum dot inside a microcavity.
\emph{Phys. Rev. B} \textbf{78}, 125318 (2008).





\bibitem{spin4} Wei, H. R. \& Deng, F. G. Universal quantum gates on
electron-spin qubits with quantum dots inside single-side optical
microcavities. \emph{Opt. Express} \textbf{22}, 593--607 (2014).




\bibitem{spin5} Wang, H. F., Zhu, A. D., Zhang, S. \& Yeon, K. H. Optically
controlled phase gate and teleportation of a controlled-NOT gate for
spin qubits in a quantum-dot-microcavity coupled system. \emph{Phys.
Rev. A} \textbf{87}, 062337 (2013).






\bibitem{NV1} Togan, E. \emph{et al.} Quantum entanglement
between an optical photon and a solid-state spin qubit.
\emph{Nature}  \textbf{466}, 730--734 (2010).


\bibitem{NV2} Wei, H. R. \& Deng, F. G. Compact quantum gates on electron-spin
qubits assisted by diamond nitrogen-vacancy centers inside cavities.
\emph{Phys. Rev. A} \textbf{88}, 042323 (2013).


\bibitem{NV3} Neumann, P. \emph{et al.} Quantum register based on coupled electron
spins in a room-temperature solid.  \emph{Nat. Phys.} \textbf{6},
249--253 (2010).




\bibitem{superconducting1}  Yamamoto, T., Pashkin, Y. A., Astafiev, O., Nakamura, Y. \&
Tsai, J. S. Demonstration of conditional gate operation using
superconducting charge qubits.  \emph{Nature} \textbf{425}, 941--944
(2003).



\bibitem{superconducting2}
Clarke, J. \& Wilhelm, F. K. Superconducting quantum bits.
\emph{Nature} \textbf{453}, 1031--1042 (2008).



\bibitem{microwave1}  Hua, M., Tao, M. J. \& Deng,  F. G. Fast universal quantum gates
on microwave photons with all-resonance operations in circuit QED.
\emph{Sci. Rep.} \textbf{5}, 9274  (2015).



\bibitem{microwave2}  Hua, M.,  Tao, M. J. \& Deng,  F. G.  Universal quantum gates on
microwave photons assisted by circuit quantum electrodynamics.
\emph{Phys. Rev. A} \textbf{90}, 012328 (2014).




\bibitem{hybrid} Bonato, C. \emph{et al.}
 CNOT and Bell-state analysis in the
weak-coupling cavity QED regime. \emph{Phys. Rev. Lett.}
\textbf{104}, 160503 (2010).



\bibitem{Wei} Wei, H. R. \& Deng, F. G. Universal quantum gates for hybrid systems assisted by quantum dots inside
double-sided optical microcavities. \emph{Phys. Rev. A} \textbf{87},
022305 (2013).







\bibitem{c20} Fortier, K. M., Kim, Y., Gibbons, M. J., Ahmadi, P. \& Chapman, M. S. Deterministic loading of individual
atoms to a high-finesse optical cavity. \emph{Phys. Rev. Lett.}
\textbf{98}, 233601 (2007).



\bibitem{c11} Cirac, J. I., Zoller, P., Kimble, H. J. \& Mabuchi, H. Quantum state transfer and entanglement distribution among
distant nodes in a quantum network. \emph{Phys. Rev. Lett.}
\textbf{78}, 3221--3224 (1997).


\bibitem{c12} Boozer, A. D., Boca, A., Miller, R., Northup, T. E. \& Kimble, H. J. Reversible state transfer between light
and a single trapped atom. \emph{Phys. Rev. Lett.} \textbf{98},
193601 (2007).


\bibitem{c13} Duan, L. M., Kuzmich, A. \& Kimble, H. J. Cavity QED and quantum-information processing with ``hot'' trapped atoms.
\emph{Phys. Rev. A }\textbf{67}, 032305 (2003).


\bibitem{c14} Duan, L. M. \& Kimble, H. J. Scalable photonic quantum computation through cavity-assisted interactions.
\emph{Phys. Rev. Lett.} \textbf{92}, 127902 (2004).


\bibitem{c15} Duan, L. M., Wang, B. \& Kimble, H. J. Robust quantum gates on neutral atoms with cavity-assisted photon scattering.
\emph{Phys. Rev. A} \textbf{72}, 032333 (2005).


\bibitem{c16} Cho, J. Y. \& Lee, H. W. Generation of atomic cluster states through the cavity input-output process.
\emph{Phys. Rev. Lett.} \textbf{95}, 160501 (2005).



\bibitem{c21} Liang, L. M. \& Li, C. Z. Realization of quantum SWAP gate between flying and stationary qubits.
\emph{Phys. Rev. A} \textbf{72}, 024303 (2005).




\bibitem{c26} Wang, B. \& Duan, L. M. Implementation scheme of controlled SWAP gates for quantum fingerprinting
and photonic quantum computation. \emph{Phys. Rev. A} \textbf{75},
050304(R) (2007).




\bibitem{c28} Lin, X. M., Xue, P., Chen, M. Y., Chen, Z. H. \& Li, X. H. Scalable preparation of multiple-particle
entangled states via the cavity input-output process. \emph{Phys.
Rev. A} \textbf{74}, 052339 (2006).





\bibitem{Reisterer1} Reiserer, A., Ritter, S. \& Rempe, G. Nondestructive detection of an optical photon.
\emph{Science} \textbf{342}, 1349 (2013).


\bibitem{Tiecke} Tiecke, T. G. \emph{et al.}
Nanophotonic quantum phase switch with a single atom.  \emph{Nature}
 \textbf{508}, 241--244 (2014).

\bibitem{Reisterer2} Reiserer, A., Kalb, N., Rempe, G. \& Ritter, S. A quantum gate between a flying optical photon and a single trapped atom.
\emph{Nature} \textbf{508}, 237--240 (2014).


\bibitem{Kalb} Kalb, N., Reiserer, A., Ritter S. \& Rempe, G. Heralded storage of a photonic quantum bit in a single atom.
\emph{Phys. Rev. Lett.} \textbf{114}, 220501 (2015).


\bibitem{Turchette} Turchette, Q. A., Hood, C. J., Lange, Q., Mabuchi, H., and Kimble, H. J., Measurement of conditional phase shifts for quantum logic.
\emph{Phys. Rev. Lett.} \textbf{75}, 25 (1995).


\bibitem{c30} Dayan, B. \emph{et al.}
A photon turnstile dynamically regulated by one atom. \emph{Science}
\textbf{319}, 1062 (2008).


\bibitem{c25} Xiao, Y. F., Han, Z. F. \& Guo G. C. Quantum computation without strict strong
coupling on a silicon chip. \emph{Phys. Rev. A} \textbf{73}, 052324
(2006).


\bibitem{An} An, J. H., Feng, M. \& Oh, C. H. Quantum-information processing with a single photon
by an input-output process with respect to low-Q cavities.
\emph{Phys. Rev. A} \textbf{79}, 032303 (2009).


\bibitem{An6} Chen, Q. \& Feng, M. Quantum gating on neutral atoms in low-Q cavities by a single-photon input-output process.
\emph{Phys. Rev. A} \textbf{79}, 064304 (2009).


\bibitem{An5} Song, J., Xia, Y. \& Song, H. S. Quantum gate operations using atomic qubits through cavity input-output process.
\emph{Eur. Phys. Lett.} \textbf{87}, 50005 (2009).



\bibitem{An1} Bastos, W. P., Cardoso, W. B., Avelar, A. T. \& Baseia, B. A note on entanglement swapping of atomic states
through the photonic Faraday rotation.  \emph{Quant. Inf. Process.}
\textbf{10}, 395--404 (2011).





\bibitem{An4} Mei, F., Yu, Y. F., Feng, X. L., Zhu, S. L. \& Zhang, Z. M. Optical quantum computation with cavities in the intermediate
coupling region. \emph{Eur. Phys. Lett.} \textbf{91}, 10001 (2010).


\bibitem{An3} Su, S. L., Guo, Q., Zhu, L., Wang, H. F. \& Zhang, S. Atomic quantum information processing in low-Q cavity in
the intermediate coupling region. \emph{J. Opt. Soc. Am. B}
\textbf{29}, 2827--2833 (2012).


\bibitem{An7} Mei, F., Feng, M., Yu, Y. F. \& Zhang, Z. M. Scalable quantum information processing with atomic ensembles and flying photons.
\emph{Phys. Rev. A} \textbf{80}, 042319 (2009).


\bibitem{An2} Bastos, W. P., Cardoso, W. B., Avelar, A. T., de Almeida, N. G. \& Baseia, B. Controlled teleportation via photonic Faraday
rotations in low-Q cavities. \emph{Quant. Inf. Process.}
\textbf{11}, 1867 (2012).


\bibitem{c32} Walls, D. F. \& Milburn, G. J. Quantum optics (Springer-Verlag, Berlin, 1994).


\bibitem{Reisterer3} Reiserer, A., N$\ddot{o}$lleke, C., Ritter, S. \& Rempe, G.
Ground-state cooling of a single atom at the center of an optical
cavity. \emph{Phys. Rev. Lett.} \textbf{110}, 223003 (2013).\\




\end{thebibliography}
\end{document}